\documentstyle[twoside,fleqn,npb,epsfig]{article}
\newcommand{\AmS}{{\protect\the\textfont2
  A\kern-.1667em\lower.5ex\hbox{M}\kern-.125emS}}

\def\mb#1{\mbox{\scriptsize {#1}}}

\title{\vspace{-1.5cm}
{\small CERN-TH/99-196\\[1pt]
        GLAS-PPE/1999-08\\[-7pt]
	hep-ex/9907011}\\[1cm]
Progress and Problems in QCD --- \\
Report from the Hadronic Final States Working Group at DIS99}

\author{Matteo Cacciari\address{CERN, Theory Division, CH-1211 
                                Geneva 23, Switzerland.},
        Frank Chlebana\address{Fermi National Accelerator Laboratory,
        P.O. Box 500, Batavia, IL 60150, U.S.A.},
        Laurel Sinclair\address{Department of Physics and Astronomy,
        Glasgow University, Glasgow G12 8QQ, UK.} and
        Marc Weber\address{ Institut f\"ur Hochenergiephysik, 
        Universt\"at Heidelberg, Schr\"oderstra\ss e 90, D--69120 
        Heidelberg, Germany.}
}

\begin{document}

\begin{abstract}
We present a summary of the Hadronic Final States parallel sessions
of the DIS99 workshop. Topics were presented over two days and included
both theoretical and experimental talks. 
Recent progress in the understanding of QCD in deep inelastic scattering, 
$e^+e^-$ collisions, and in $\gamma$ and p collisions was discussed.
\end{abstract}

\maketitle

\section{THEORY SUMMARY}

During the parallel sessions of the Hadronic Final States Working Group about
forty talks were presented. Out of these, about ten might be classified as
theoretical ones. The issues they discussed ranged from instantons to heavy
flavour production to higher order QCD calculations. It is therefore
understandable how difficult it might be to try to provide an organic summary
of such a widespread collection of interesting topics.

We shall therefore try to highlight here what we consider to be the main
points of the various contributions, leaving the task of properly 
introducing and explaining the subject to the individual summaries. Such
summaries will  not be  cited explicitly in this section, since they can  easily 
be found in
these proceedings under  the   name of the  person who gave the presentation.

A common feature can be identified in almost all the talks given. They clearly
focus on the need to go beyond standard fixed next-to-leading order (NLO) QCD
perturbation theory as the quality of the experimental data demands more
accurate theoretical calculations.

G. Salam and H. Jung both described phenomenological studies related to the
CCFM equation. This is an evolution equation that goes beyond the so-called
``multi-Regge'' limit of the BFKL equation, and also tries to implement effects
due to colour coherence (via angular ordering) and soft particles. 
As CCFM is
harder to solve than BFKL, the question is how similar the predictions of 
the two approaches are. 
Salam argued that, with the inclusion of soft effects, BFKL
and CCFM can be shown to lead to identical predictions at the leading log
level. Differences at sub-leading level, as well as ambiguities in the
implementation of the equations at this level, do however remain. Jung
described a practical implementation of the CCFM equation in the
program SMALLX and showed that a good phenomenological description of both
$F_2$ and forward jets data can be achieved. 

Small-$x$ logarithms  are  not the only ones appearing in a QCD perturbative
expansion. N.~Kidonakis described how to take care of the large logs resulting
from soft gluon emissions when final states are produced   near threshold.
Resummed expressions for these terms have been written some time ago, but until
recently it was still unclear how to treat these expressions so that the final
result  would indeed only contain the resummation  of the perturbative series 
and  not further  spurious  effects. In the approach that Kidonakis described,
the resummed expression is truncated at  next-to-next-to-leading order, i.e.
one order beyond what is available as a full fixed order calculation.  Due to
the fairly good  convergence properties  of the series, such  a treatment
suffices to bring an improvement over the standard NLO
calculation, and at the
same time ensures consistency with the  perturbative expansion.

One further resummation issue was addressed by S. Kretzer who described how
charm effects in parton  distribution functions, appearing at fixed order as
$\log(Q/m)$, can be resummed using the ACOT scheme.
He studied both  neutral and charged current deep inelastic scattering (DIS).
The 
difference   between fixed order  and  resummed  predictions  does not seem to
be within  the reach  of present experimental  accuracy. It  is  however likely
that   such resummed approaches will be more and more important in the future,
as experimental precision improves  and larger scales  are probed.

S. Frixione  reviewed the status of theoretical calculations of heavy
quark production. At small and
moderate transverse  momenta, NLO QCD calculations are available  and reliable.
In the large transverse momentum  region, on the other hand, large logs of the
form $\log(p_T/m)$ develop. Resummation techniques for such terms have been 
developed in
recent years. By making  use of the Altarelli-Parisi evolution of a
perturbatively calculable heavy quark fragmentation function these logs are
resummed to all orders. Inclusion of a non-perturbative parametrization for
heavy quark hadronization effects, such as  the Peterson fragmentation
function,   also provides a description of $D^*$
photoproduction data. Frixione  did however warn that this resummed
calculation should not be used at too small transverse  momentum values; a
proper combination with the full fixed-order NLO massive calculation is
necessary to  make  it reliable  in this region.

A more exotic feature of QCD, namely instantons, which also go beyond
perturbation theory, was discussed  by F. Schrempp. Such objects, originating
from the rich structure of the QCD vacuum, can   in principle play an important
role in various long distance aspects of QCD. There are also
short-distance implications. It was argued in this talk that  in DIS  
at HERA there exists a potential for observing instanton-induced
processes otherwise  forbidden by usual QCD  perturbation theory.

Back on more standard ground, D. Soper and  M. Grazzini described efforts to
produce calculations in fixed order perturbation theory. Soper described  a
numerical approach to one-loop computations: rather than evaluating each
diagram analytically, all diagrams are properly added and the integrals
involved are performed numerically, in such a way that the  individual
singularities cancel before integrating. The final result is therefore finite.
This method has so far been successfully applied to the evaluation of the thrust
distribution in $e^+e^-$ collisions, but can in principle be extended to other
processes. Grazzini reported  on progress in perturbative calculations
beyond the one-loop approximation. In particular, he described studies of the 
collinear limits of three partons, where generalizations of the
Altarelli-Parisi splitting vertices appear. Such limits are one  of the
building blocks for next-to-next-to-leading calculations, the next frontier of
perturbative QCD.

\section{HEAVY QUARK PRODUCTION}

Experimental results, both on charm and on bottom production, were presented by
the ZEUS and H1 collaborations. Many $D^*$ production data were updated from
previous measurements. 
Heavy quarks can provide particularly
interesting tests of perturbative  QCD since, due to their large mass acting as
a cutoff for infrared singularities, their total production rate can be
predicted on rigorous theoretical grounds with no  free  parameters other than
their mass, the parton distribution functions and the strong coupling.

\begin{figure}[htb]
\centerline{\epsfig{file=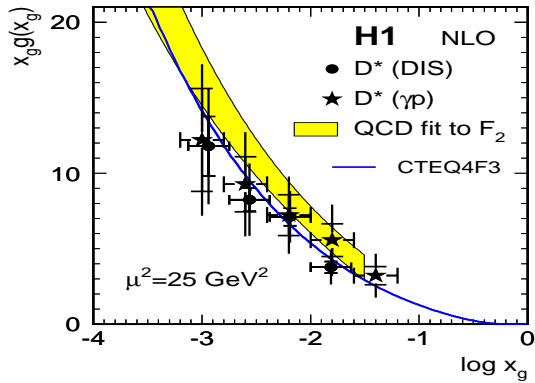,height=5cm,width=7cm}}
\vspace{-1.cm}
\caption{Gluon densities obtained from $D^*$ analyses in photoproduction
and DIS.}
\label{fig:H1_glue_fromD}
\end{figure}

The H1 collaboration showed that a
generally good, albeit not perfect, agreement with NLO QCD predictions can be
observed~\cite{R.Gerhards}. It is also worth noticing that the gluon distribution function in the
proton has been extracted from charm data in both photoproduction and DIS.  
The two results show good agreement with each  other  
and with the
indirect determination from $F_2$, as shown in Fig.~\ref{fig:H1_glue_fromD}.

\begin{figure}[htb]
\centerline{\epsfig{file=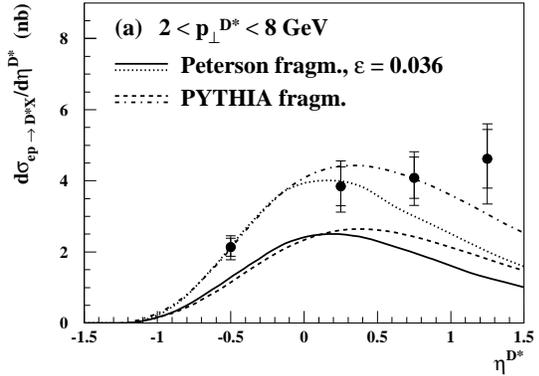,height=5cm,width=7cm}}
\vspace{-1.cm}
\caption{ZEUS differential cross sections $d\sigma / d\eta^{D^*}$ compared with
NLO predictions for the massive approach.  Full description of curves
may be found in~\cite{Y.Eisenberg}.}
\label{fig:ZEUS_D}
\end{figure}

ZEUS $D^*$ data were  shown  in a new low photon-proton energy region,
80~GeV~$<W_{\gamma p}<120$~GeV~\cite{Y.Eisenberg}. 
As shown in Fig.~\ref{fig:ZEUS_D}, the experimental  data tend to  be 
generally higher than  the NLO QCD predictions, 
especially in the forward (proton) direction. Comparisons
with  the so-called massless approach show better agreement, but  this kind of
approach is probably  not fully reliable at these low transverse momentum
values~\cite{S.Frixione}.

Bottom production data, given in terms of  observed leptons coming from
semileptonic decays, were also presented by both ZEUS and H1 collaborations.
ZEUS observed electrons, finding a cross section larger by a factor of $\sim 4$
than the one  predicted by the 
HERWIG leading order (LO) plus parton shower Monte Carlo program~\cite{M.Wing}. 
H1 performed two analyses,
both using muons  to tag the heavy quark~\cite{P.Newman}. In these  cases,  the
experimental  results are about  a factor of 5 and  3 
larger than the  prediction of a different LO Monte Carlo program, AROMA. 

Such large factors between   theory and experiment might seem 
important. However, one should keep in mind both the large errors still present
in the experimental determinations  and the fact  that   the theoretical
prediction is given here by a leading order Monte Carlo program.  It is well known 
that NLO calculations  for heavy quarks  greatly  increase leading  order
predictions,  usually by at least a factor of 2. However, a comparison 
with  a NLO
calculation  is unfortunately  not  yet possible,  since experimental
results are only quoted for a ``visible'' cross  section for which no full
simulation at NLO is so  far available.
Combining this fact with the
experimental  uncertainty,  it becomes  apparent that it is still premature  to
talk of  a discrepancy in the bottom  production  cross section at HERA. We are
of course   looking  forward to  more detailed comparisons between  theory and
experiment, bottom production being better behaved   in QCD perturbation theory
than  charm,  and therefore providing a better test of QCD predictions.

Heavy quark production  was also discussed for  the hidden flavour case, i.e.
heavy quarkonia  production.
Elastic and inelastic
$J\!/\!\psi$  electroproduction  has been analysed   by the H1 
collaboration~\cite{S.Mohrdieck}. 
Results from an  inclusive sample  (elastic plus inelastic) were  compared to
predictions from the  so-called soft colour interaction  model  as  implemented
in the  Monte Carlo  program AROMA. Most  of the shapes of the differential
distributions can generally be reproduced,  while the  magnitude of the
theoretical predictions is usually too low. Results from an inelastic sample
were instead  compared to  predictions  obtained with the non-relativistic QCD
(NRQCD)
approach to  quarkonium production.  Such predictions can be   seen to be at 
some variance with the data,  both  in magnitude and, especially for the
rapidity distribution, in shape. It is however  known that,  especially at such
a low scale as the one set by the charm mass,  NRQCD predictions   can suffer
from large uncertainties.

\section{JETS IN DIS}

The use of jet algorithms and the study of jet-related 
observables continue to be both a popular and powerful approach
in order to characterize the hadronic final state's properties.
The analysis of multi-jet events in DIS has been used to extract 
the value of the strong coupling constant $\alpha_s$ 
\cite{alphas1,alphas2,alphas3,alphas4}. These and many other important 
analyses have clearly shown the large 
potential of QCD studies with jets at HERA but have also revealed 
important limitations. The latter are caused by the difficulties of 
current QCD Monte Carlo models to describe the data precisely, by the 
large renormalization scale uncertainties of QCD predictions in NLO in 
the phase space covered, or by the uncertainties of the proton's 
parton density. 

In this session various presentations of jet analyses were given, 
which addressed these issues. Most analyses profited from the 
large data sample collected in particular during the very successful 
'97 data taking period of HERA. Thus, generally the measurements were 
considerably extended, either into the region of high photon virtuality, $Q^2$, 
or to 
harder jet structures corresponding to large transverse jet energies,
$E_T^{Breit}$, in the Breit frame. Three representative 
jet analyses are discussed in more detail below.

\begin{figure}[h!]
\centerline{\epsfig{file=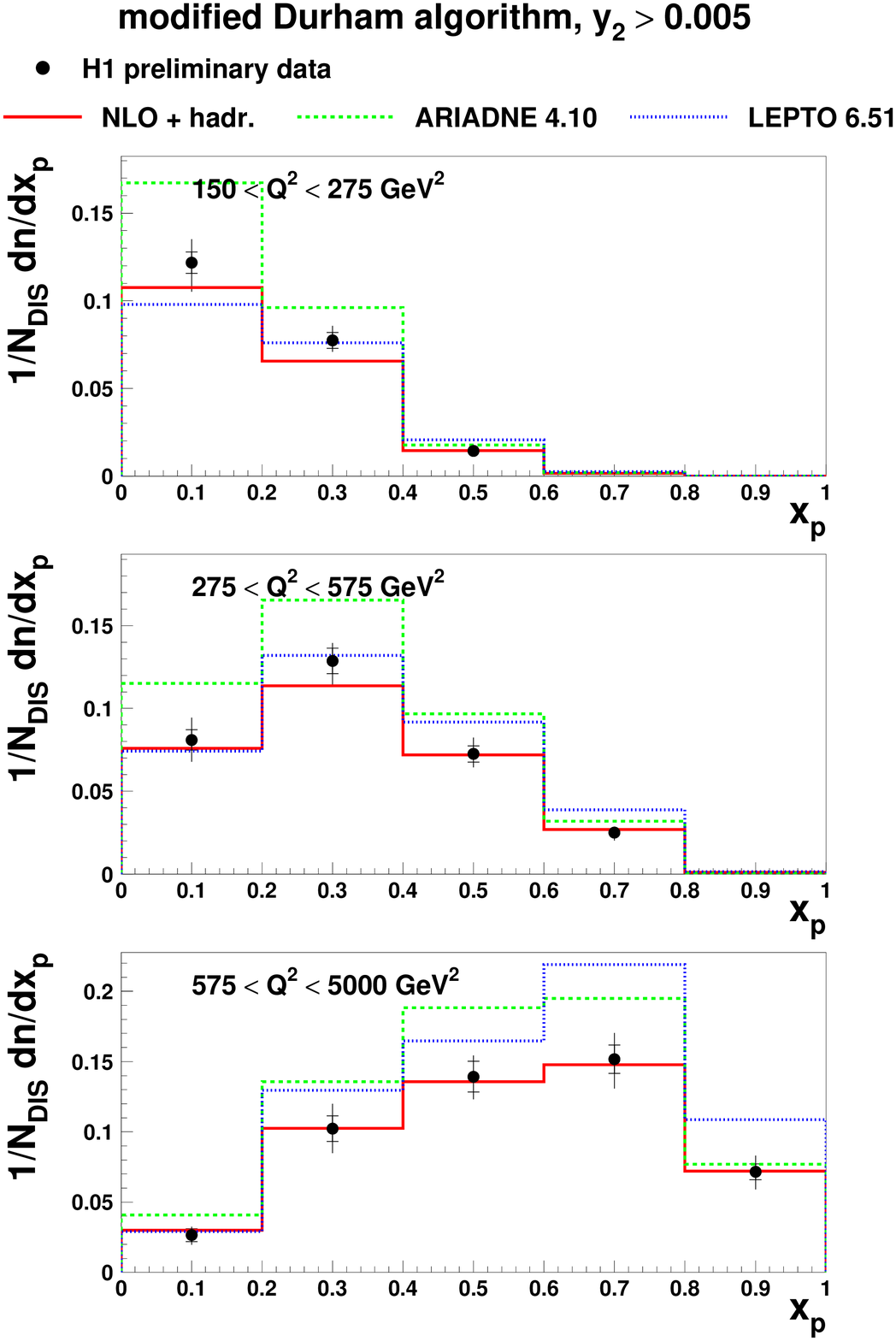,height=9cm,width=7.3cm}}
\vspace{-1.cm} 
\caption{Distribution of $x_p = x_{\mb{Bj}}/\xi$ in three ranges of $Q^2$ 
measured with the modified Durham algorithm.}
\label{fig1mw}
\end{figure}

In \cite{N.Tobien} a systematic comparison of various measured jet 
distributions with model predictions was performed in dijet events at 
$Q^2 > 150$ GeV$^2$. 
In Fig.~\ref{fig1mw} the $x_p$ distribution determined with the 
modified Durham algorithm (run in the laboratory frame) is shown for three 
different ranges of $Q^2$. The precision of the data is high and significant 
deviations of the QCD Monte Carlo models ARIADNE and LEPTO are observed. 
In contrast to the MC models, the perturbative QCD predictions in NLO, which 
are also shown in Fig.~\ref{fig1mw}, describe the measured jet 
distributions very well. 
The same conclusions are reached using the factorizable $k_T$ algorithm 
(run in the Breit frame).

It is important to note that both Monte Carlo models have recently been 
investigated in the context of the HERA Monte Carlo Workshop \cite{MCworkshop} 
and were considerably improved. In version 4.10 of ARIADNE, the $Q^2$ 
dependence of dijet events has been modified. In version 6.51 of LEPTO, a 
new model of soft-colour interactions has been implemented. 
Given the relative disagreement with the data, further development of the 
models is desirable. 

An excellent description of the data by QCD in NLO is also 
observed in the preliminary inclusive jet cross sections, 
$d^2\sigma_{jet}/dE_{T}dQ^{2}$, in the kinematic region of 
$Q^2 > 150 $ GeV$^2$ and for transverse jet energy $E_T^{Breit} > 7$ 
GeV \cite{M.Wobisch}. In this phase space region both perturbative and 
non-perturbative uncertainties are expected to be small. 

The corresponding dijet cross section measurements were fitted 
simultaneously with $F_2$ 
determinations to yield the gluon density of the proton \cite{gluonmw}. This 
procedure extends the accessible range 
of the gluon momentum fraction, $\xi$, to larger values with respect to 
the less 
direct determinations of the gluon density from the scaling violation 
of $F_2$. The 
value of the strong coupling constant was assumed to be known in this 
analysis, but a combined fit of both the gluon density and $\alpha_s$ should 
be possible in the future.

A preliminary measurement of the dijet event rate and the dijet 
cross section as a function of $Q^2$ and a determination of $\alpha_s$ 
were presented in \cite{E.Tassi}. Jets of high transverse momentum 
$E_T^{Breit}$ have been selected at $Q^2 > 470$ GeV$^2$. In the region of 
large 
$Q^2$, corresponding to large parton momentum fractions, the proportion of 
gluon-initiated scattering processes is minimal. Thus this analysis 
is less sensitive to the uncertainties in the gluon density and may 
depend to a lesser extent on the details of parton density extractions 
from the world data than earlier $\alpha_s$ determinations. Also, the 
renormalization scale uncertainties are (relatively) small, both owing to the 
large value of $Q^2$ and the high jet transverse energies required. Again, 
this nicely illustrates the quest for smaller (theoretical) uncertainties 
by selecting more restrictive but safer phase space regions.


\begin{figure}[h!]
\centerline{\epsfig{file=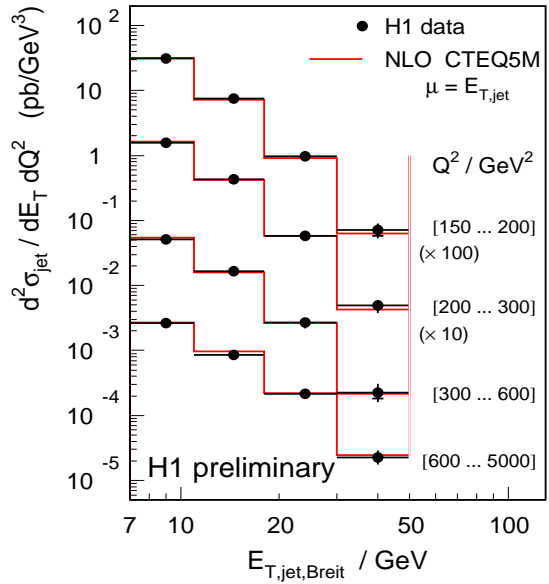,height=8cm,width=7.5cm}}
\vspace{-1.cm} 
\caption{Inclusive jet cross section as a function of the 
transverse jet energy $E_T$ in the Breit frame in different 
regions of $Q^2$.}
\label{fig2mw}
\end{figure}

\section{EVENT SHAPES AND POWER CORRECTIONS}

The comparison of exclusive hadronic final state observables 
with the corresponding perturbative QCD predictions requires 
an estimation of hadronization effects. Phenomenological 
models, such as the string or the cluster model implemented in Monte Carlo 
generators, are primarily used for these estimates. Depending on the observable 
under study, hadronization corrections may be large, however, and it 
is difficult to estimate their uncertainty rigorously. 

An alternative approach consists in applying analytical power corrections 
of the form $\sim$ 1/$Q^p$ when comparing  
perturbative QCD predictions with measured hadronic distributions. 
The leading power $p$ and the exact form 
of the power corrections to the mean values of event shape 
distributions have been calculated for a large number of 
observables in both $e^+e^-$ annihilation and deep inelastic 
scattering. The size of the 
correction depends on the value of an effective universal coupling constant  
$\overline{\alpha}_0$, on the strong coupling constant $\alpha_s$ and, 
of course, on $Q^2$. 

\begin{figure}[h]
\centerline{\epsfig{file=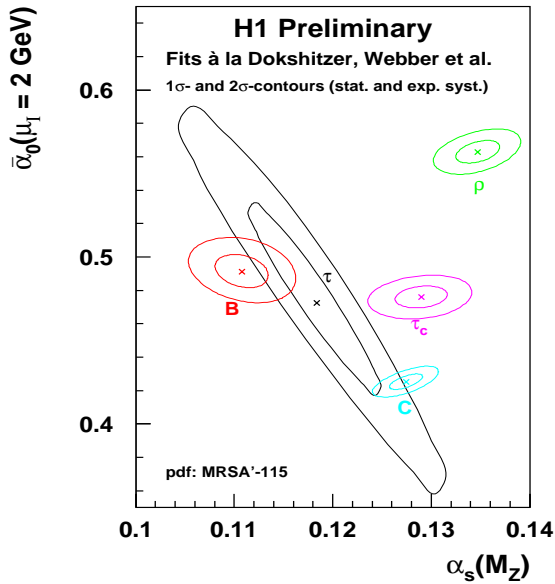,height=8cm,width=7.5cm}}
\vspace{-1.cm} 
\caption{Determinations of $\alpha_s(M_Z)$ and $\overline{\alpha}_0$
for various event shape variables measured in DIS.}
\label{fig3mw}
\end{figure}

The results of a 2-parameter fit of $\alpha_s$ and $\overline{\alpha}_0$
to the mean values of various event shape distributions recently measured 
at HERA~\cite{K.Rabbertz} are shown in Fig.~\ref{fig3mw}. These results 
consider the recent theoretical developments of the calculation of two-loop 
corrections, leading to the so-called Milan factor~\cite{Milan}, and of an 
updated calculation of the coefficient for the jet 
broadening variable~\cite{jetbroad}. Also, the experimental precision 
has been considerably improved with respect to an earlier 
analysis~\cite{evshapH1}. The value of $\overline{\alpha}_0$ is 
found to be $\sim 0.5$ and it is independent of the variable studied to within 
20$\%$. The variation of the resulting $\alpha_s$ values is very large, 
however,  which strongly suggests that further theoretical studies are needed.

\begin{figure}[h]
\epsfig{file=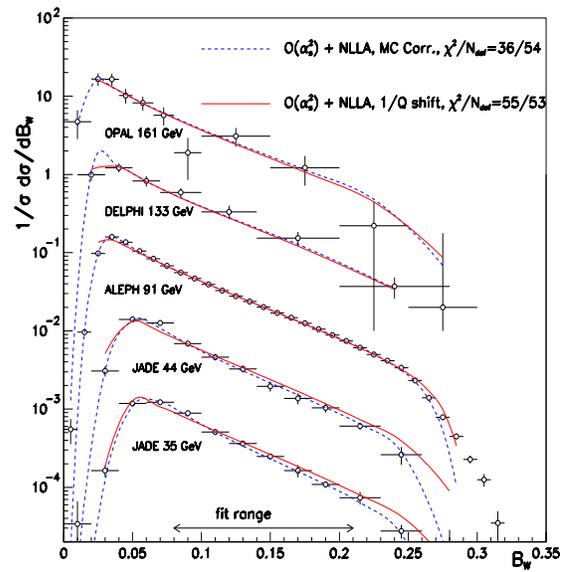,height=8cm,width=7.5cm}
\vspace{-1.cm} 
\caption{Distributions of the wide jet broadening, $B_W$, at different 
centre-of-mass energies together with perturbative QCD predictions 
combined with power corrections or traditional hadronization corrections.}
\label{fig4mw}
\end{figure}

In $e^+e^-$ annihilation similar fits to the mean values (as a function of 
the centre-of-mass energy $\sqrt{s}$) of thrust, wide jet broadening 
and heavy jet mass yielded much better agreement in $\alpha_s$ and 
similar agreement in $\alpha_0$ \cite{G.Dissertori}. Predictions 
of power corrections to distributions, in contrast to mean values, of 
event shape observables have also been tested \cite{G.Dissertori}.
Fig.~\ref{fig4mw} shows the wide jet broadening distribution measured at 
different centre-of-mass energies. Perturbative QCD predictions with 
either hadronization corrections from Monte Carlo models 
(full line) or power corrections (dashed line) are fitted to the 
data and yield fits of similar quality to those without the power corrections
but systematically lower values of $\alpha_s$. The differences are largest 
at low centre-of-mass energy where also the applied corrections 
are sizeable.

In conclusion, the concept of power corrections has led to a remarkable 
theoretical activity, often in close collaboration with experimentalists, and 
power corrections prove to be surprisingly successful. The approach is rather 
economical in the sense that only one additional universal parameter 
$\alpha_0$ needs to be determined by experiment. A number of  
examples have  recently been given, where also the limitations and 
difficulties became  apparent. 

\section{FRAGMENTATION IN DIS}

In the above jet analyses, the emphasis is placed on accessing the very rare 
multi-jet events at increasingly large jet $E_T$ to minimize 
soft non-perturbative effects in the subsequent comparison to perturbative 
QCD predictions. Valuable information on the properties of QCD can also be 
gained from the study of charged particle production, however, which is 
obviously strongly influenced by the non-perturbative hadronization phase. 
A key question is: To 
what extent do the measured hadrons reflect the underlying parton spectra?
These depend on the initial partonic configuration and the subsequent 
parton cascade. The typical observables are particle rates, momentum 
distributions of the hadronic final state particles and multi-particle 
correlation variables. Increasingly refined perturbative predictions for more 
and more complex observables have been derived within the framework of the 
Modified Leading Log Approximation (MLLA) \cite{MLLA}. Many of those 
predictions have 
been compared with data using the concept of Local Parton-Hadron Duality 
(LPHD). The hypothesis of LPHD in connection with MLLA relates the average 
properties of partons and hadrons by means of a simple normalization constant.
 With these assumptions, the theoretical predictions depend essentially on 
two parameters only, an effective strong coupling constant determined by the 
QCD scale $\Lambda$ and an energy cutoff parameter $Q_0$. Both have to be 
determined by measurement.

\begin{figure}[b!]
\centerline{\epsfig{file=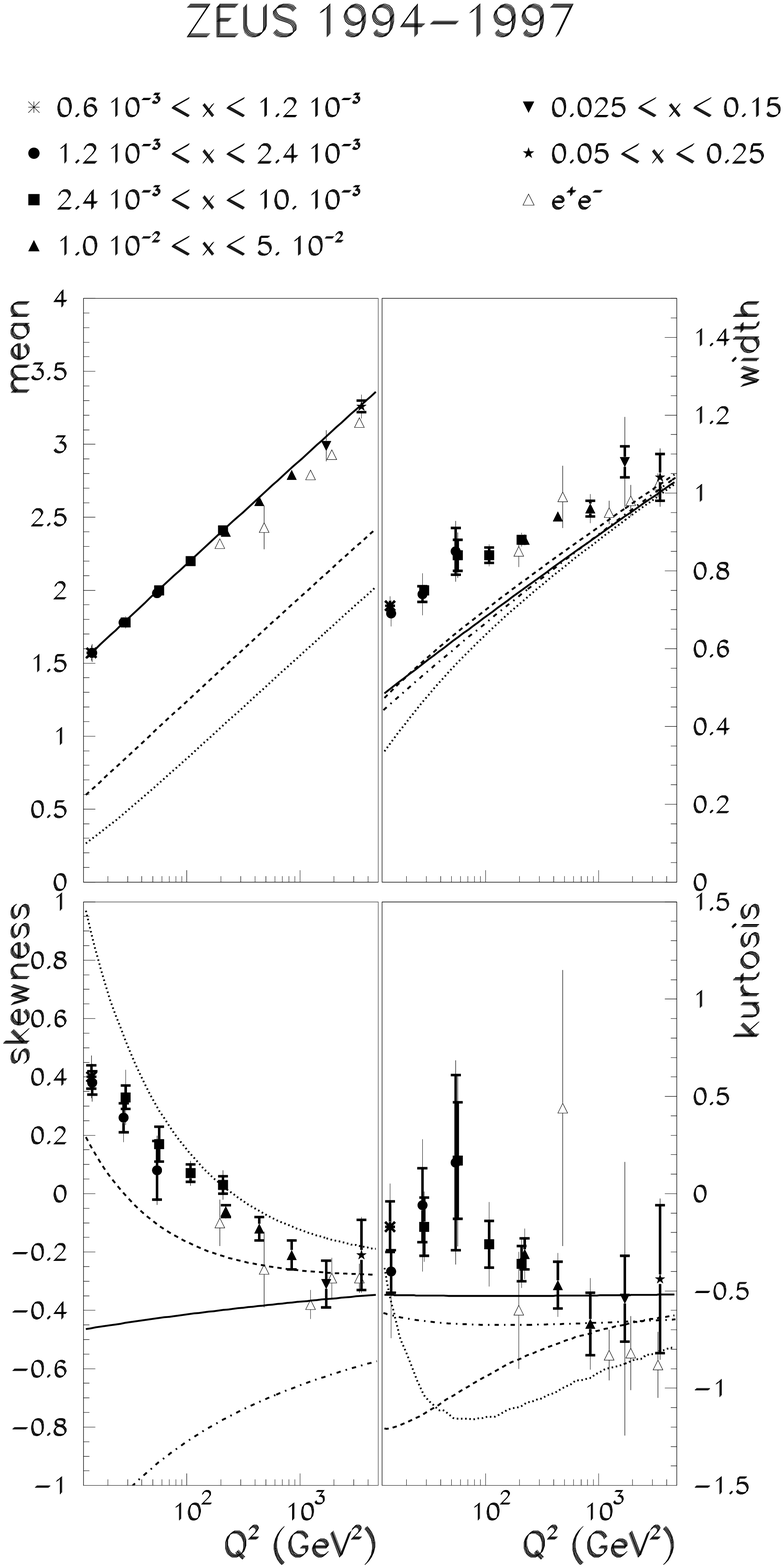,height=9cm,width=8cm}}
\vspace{-1.cm} 
\caption{Evolution of the mean, width, and of the higher moments 
skewness and kurtosis, of the $\ln(1/x_p)$ distribution with $Q^2$.}
\label{fig5mw}
\end{figure}

A recent measurement of scaled momentum distributions~\cite{L.Zawieski} 
is presented in 
Fig.~\ref{fig5mw},  where the  mean value and higher moments of the $\xi$ 
distribution are shown as a function of $Q^2$ for different ranges of $x$. 
The variable $\xi$ is defined as $\ln 1/x_p = \ln Q/2p^{Breit}$. Only 
particles in the current hemisphere of the Breit frame are considered. 
This makes possible a direct comparison with results from $e^+e^-$ 
annihilation, which are also included in the figure. 

The fair agreement observed between the results of DIS and $e^+e^-$ annihilation 
 suggests that the main features of quark fragmentation 
are universal. While the mean value of the $\xi$ distribution as 
a function of $Q^2$ is well described by MLLA predictions, no consistent 
description of mean, width, skewness and kurtosis can be achieved.
The measurements benefit considerably from the large range in $Q^2$ that 
is now covered by HERA. Clearly, further studies, possibly considering 
mass effects~\cite{Ochs}, are needed to understand this disagreement.


Measurements of the properties of particles in the target hemisphere have 
also been made and the correlation between the mean particle multiplicity 
in the two hemispheres has been studied \cite{L.Zawieski}. 

\begin{figure}[h!]
\centerline{\epsfig{file=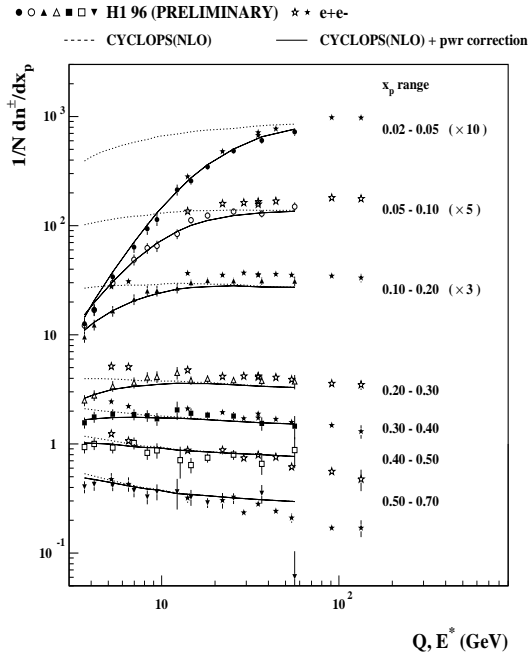,height=9.25cm,width=7.5cm}}
\vspace{-1.cm} 
\caption{The inclusive charged particle distributions 
$1/N_{DIS} dn^{\pm}/dx_p$ in the current hemisphere of the Breit frame. The 
distribution is shown as a function of $Q$ for different intervals of $x_p$. 
Data from $e^+e^-$ annihilation as a function of the centre-of-mass 
energy $E^*$ are also shown.}
\label{fig6mw}
\end{figure}

A measurement of the charged particle $x_p$ distribution as a function of 
$Q$ is shown in Fig.~\ref{fig6mw} for different ranges in 
$x_p$~\cite{D.Kant}. 
QCD calculations in NLO combined with NLO fragmentation functions are 
compared to the measurement. For large values of $x_p$ and with 
increasing $Q$, a significant decrease of the distributions is observed, 
which is expected from scaling violations in the fragmentation functions due 
to gluon radiation. The agreement with the prediction is good, except 
at small values of $x_p$ where mass effects, which are 
not considered in the calculations, are important. 
A simple power correction ansatz as 
proposed in \cite{Durham} results in a surprisingly close description of 
the data. Detailed calculations of these power corrections have been started 
and were discussed at this meeting \cite{M.Dasgupta}.

\section{FORWARD JET AND $\pi^0$ PRODUCTION IN DIS}

The study of jet and particle production in the very forward (proton) 
region in DIS is largely motivated by the interest in the parton dynamics 
at small values of $x$. In particular, one would like to access a region 
where the validity of the BFKL equations could be tested.

Experimentally, measurements close to the edge of the detectors are 
challenging, but both  
H1 and ZEUS~\cite{fwdjets} have nevertheless succeeded to measure 
forward jet cross sections. They show a significant rise of the forward jet 
cross section with decreasing values of $x$. The measurements are in 
striking disagreement with the predictions of Monte Carlo models based on 
the traditional DGLAP parton showers. Furthermore, QCD predictions in NLO 
disagree with the data \cite{E.Mirkes}. 

Recently, several ways to describe the measurement have been found 
and were presented at this meeting~\cite{B.Poetter,A.Martin,H.Jung}.
These are: the inclusion of a resolved {\it virtual} photon contribution 
in the NLO calculations performed by \cite{B.Poetter}; the inclusion of 
a resolved {\it virtual} photon component in the Monte Carlo model 
RAPGAP~\cite{fwdjets}; and the inclusion of NLO effects in a recent BFKL 
calculation by applying higher order consistency conditions~\cite{A.Martin}.
In addition, \cite{H.Jung} obtained a good description of forward jets 
(and of $F_2$) with a modified version of the Monte Carlo program SMALLX 
based on the CCFM parton evolution equation.

\begin{figure}[h!]
\centerline{\epsfig{file=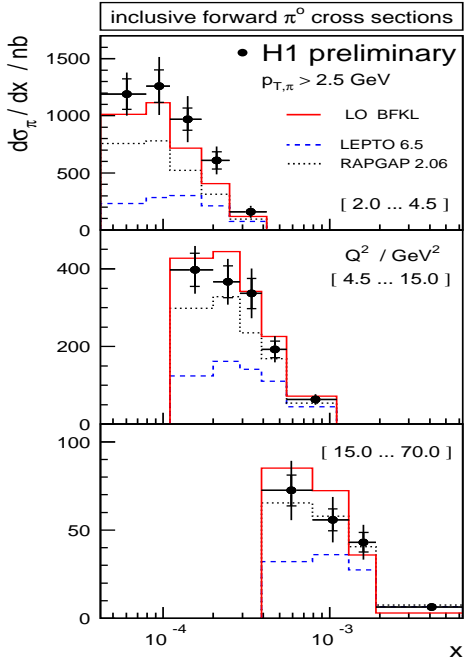,height=9cm,width=6.5cm}}
\vspace{-1.cm} 
\caption{The inclusive $\pi^0$ cross sections as a function of $x$ for 
three ranges of $Q^2$, together with the predictions of MC models and the 
BFKL calculation of \cite{A.Martin}.}
\label{fig7mw}
\end{figure}

An important new measurement of forward $\pi^0$ cross sections 
was presented in \cite{T.Wengler}. Differential distributions 
in $x$, $\eta_{\pi}$ and 
$p_{t,\pi}$ have been measured for three ranges of $Q^2$. Given 
the difficult phase space region, the precision of the measurement is 
excellent and compares favourably with the jet measurements. 
The inclusive $\pi_0$ cross sections as a function of $x$ are shown in 
Fig. \ref{fig7mw}.
Again, RAPGAP with a resolved virtual photon contribution 
describes the data better than MC models based on DGLAP parton showers 
such as  LEPTO (shown). The best description is obtained by the modified BFKL 
calculation mentioned above.

In conclusion, significant theoretical progress has been made in
understanding the physics of the forward region. The precise measurement 
of the multi-differential $\pi^0$ distributions will constrain existing 
and future models considerably.

\section{REAL PHOTON STRUCTURE}

In a hard collision involving an incoming photon, this photon may scatter
directly, or it may first fluctuate into a hadronic object.  Although it
is no longer considered possible to make a complete prediction of the 
photon's structure without input from experiment, measurements that are 
sensitive
to the hadronic nature of the photon can nevertheless provide a test of
some fundamental hypotheses.

\begin{figure}[b!]
\centerline{\epsfig{file=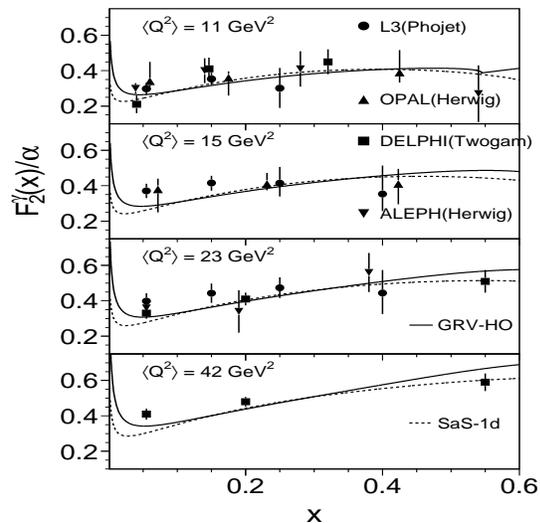,height=7cm,width=7.0cm}}
\vspace{-1.cm}
\caption{$F_2^\gamma / \alpha$ at medium $Q^2$ from LEP.}
\label{fig:f2gam_med}
\end{figure}

In $e^+e^-$ collisions, the photon's structure function $F_2^\gamma$ is
probed in deep inelastic scattering processes where one of the leptons is
scattered through a large angle.  $F_2^\gamma$ is expressed as a function
of the fraction of the photon's energy participating in the scatter, $x$,
at the resolution scale provided by the square of the momentum transfer at 
the scattered
lepton vertex, $Q^2$.  Fig.~\ref{fig:f2gam_med} shows the latest world
measurements of $F_2^\gamma(x)/\alpha$ at medium $Q^2$~\cite{K.Freudenreich}.
The available photon
parton density parametrizations are generally able to describe the data.

\begin{figure}[h!]
\centerline{\epsfig{file=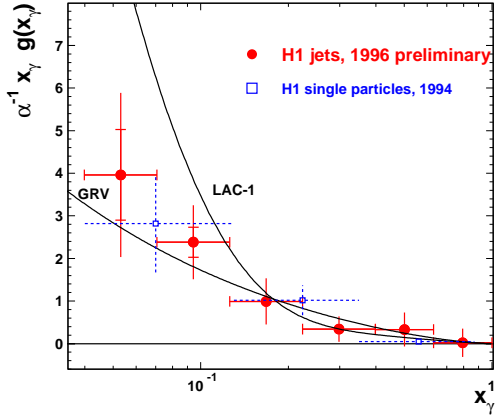,height=5.5cm,width=6.5cm}}
\vspace{-1.cm}
\caption{Measurement of the gluon density in the real photon 
from dijet and high $p_T$ track events.  }
\label{fig:H1_glue}
\end{figure}

At HERA, the partons of the proton probe directly, not only the quark 
density, but also the gluon density of the photon.  Here the resolution
scale of the probe is measured in terms of the transverse momenta of the
jets or tracks produced.  In 
Fig.~\ref{fig:H1_glue} an extraction of the leading order gluon density
of the photon is shown~\cite{J.Cvach}.
Here the probing resolution for the jet analysis
is $\langle P_T^2\rangle =74$~GeV$^2$ and for the track analysis  $\langle 
P_T^2\rangle =38$~GeV$^2$.
The different experimental approaches both indicate a rise of the leading
order gluon density as $x_\gamma$ falls.

\begin{figure}[h!]
\centerline{\epsfig{file=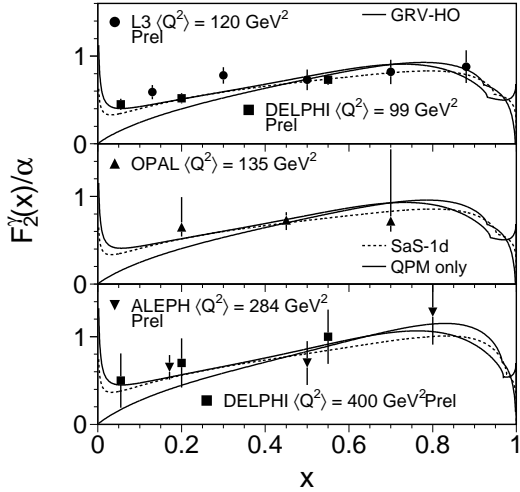,height=7cm,width=7cm}}
\vspace{-1.cm}
\caption{$F_2^\gamma / \alpha$ at high $Q^2$ from LEP.}
\label{fig:f2gam_hi}
\end{figure}

At high values of the resolution power it becomes possible to describe 
the high-$x$ quark component of the photon's structure using quark parton
model predictions alone, as shown in 
Fig.~\ref{fig:f2gam_hi}~\cite{K.Freudenreich}.
The experimental constraint on the photon's structure coming from the 
$e^+e^-$ experiments, however, grows weak here.

\begin{figure}[b!]
\centerline{\epsfig{file=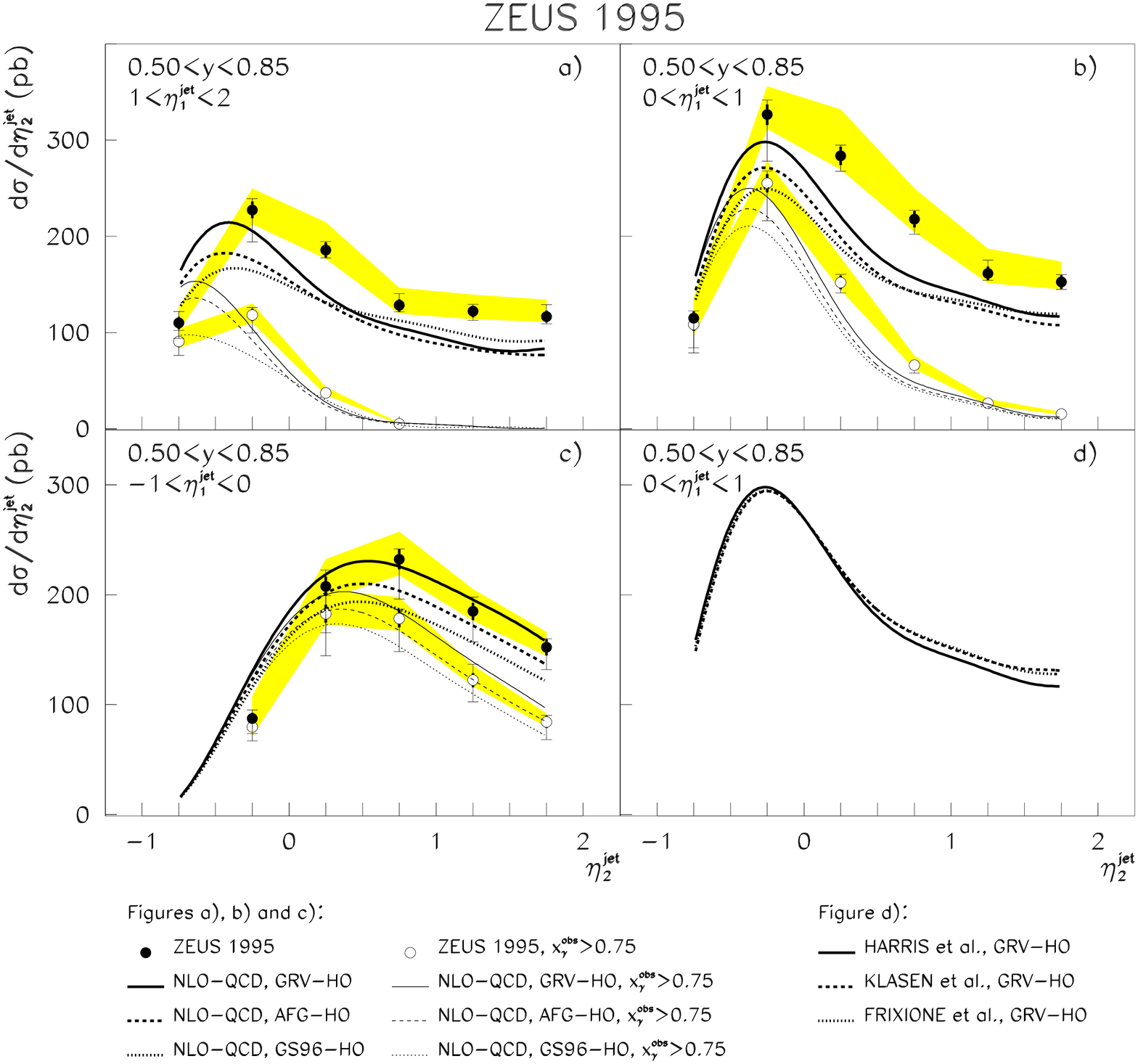,height=7cm,width=7cm}}
\vspace{-1.cm}
\caption{Dijet cross section as a function of $\eta_2^{\mb{jet}}$ in
bins of $\eta_1^{\mb{jet}}$ and for $0.50 < y < 0.85$.}
\label{fig:ZEUS_dij}
\end{figure}

ZEUS has measured dijet cross sections in photoproduction in this
high scale region 
($E_{T \mb{leading, second}}^{\mb{jet}} > 14, 11$~GeV) as shown in
Fig.~\ref{fig:ZEUS_dij}~\cite{N.Macdonald}.
Next-to-leading order perturbative QCD calculations using current
photon parton densities are unable to describe these data when
both jets are in the central region, $0 < \eta_{\mb{ 1,2}}^{\mb{jet}} < 1$.
As other uncertainties are expected to be low in this kinematic regime, it 
would be a good test of our understanding of photon-induced processes
to check whether a parton density can be found which allows the ZEUS data
to be described while remaining consistent with the available
high-$Q^2$ $F_2^\gamma$ measurements from LEP.

An interesting alternate process, which may reveal information on the
photon's structure, comes from prompt photon production in $\gamma \gamma$
and $\gamma p$ collisions.  Prompt photon cross sections have the 
potential to be free of theoretical uncertainties arising from hadronization
effects.  Both the ZEUS~\cite{K.Umemori} and TOPAZ~\cite{S.Soeldner-Rembold}
collaborations presented early studies of prompt photon production but 
a considerable improvement in statistics is necessary before any strong 
conclusions may be drawn.

\section{VIRTUAL PHOTON STRUCTURE}

It is expected that as a photon's virtuality increases, it will begin to 
lack the time to develop a complex hadronic structure.

From the total cross section for the double-tag process,
$e^+ e^- \rightarrow e^+ e^- \gamma^* \gamma^* \rightarrow e^+ e^- X$,
there is an indication that the hadronic component of the photon's structure
is still evident at sizeable photon virtualities~\cite{V.Andreev}.

\begin{figure}[h!]
\epsfig{file=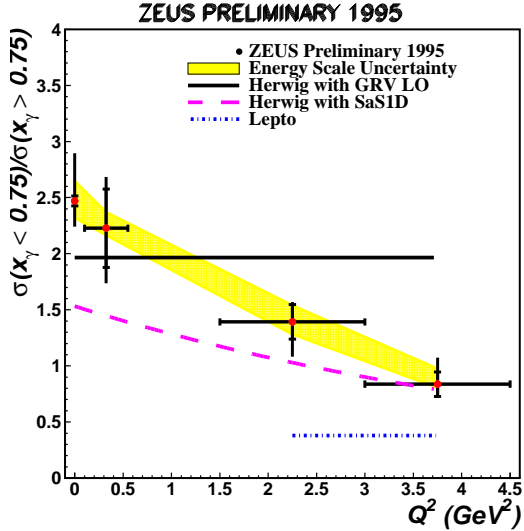,height=7.cm}
\vspace{-1.cm}
\caption{Ratio of the dijet cross section for $x_\gamma^{\mb{OBS}} < 0.75$
to the dijet cross section for $x_\gamma^{\mb{OBS}} > 0.75$.}
\label{fig:ZEUS_ratio}
\end{figure}

ZEUS has measured the ratio of the low-$x_\gamma$ dijet cross section to the
high-$x_\gamma$ dijet cross section as a function of the photon's virtuality
$Q^2$ at the probing scale provided by 
$E_T^{\mb{jets}} > 5.5$~GeV, Fig.~\ref{fig:ZEUS_ratio}~\cite{N.Macdonald}.
This ratio is flat for a parton density that does not evolve with $Q^2$ 
(GRV LO) and falling for a parton density that is suppressed with $Q^2$
(SaS 1D).  Therefore the fall that is observed in the data indicates that 
the photon's parton density is suppressed as the photon's virtuality increases.

\begin{figure}[h!]
\epsfig{file=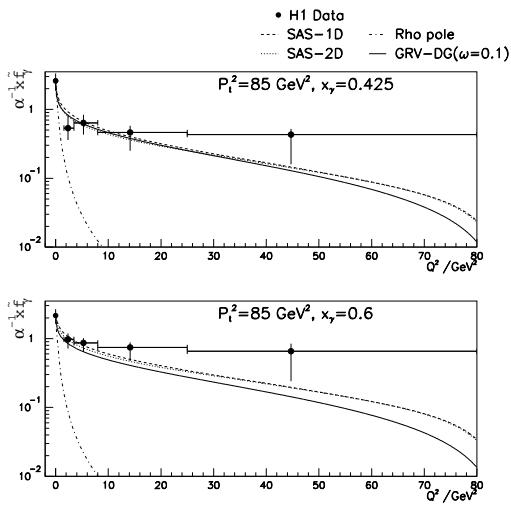,height=7.cm}
\vspace{-1.cm}
\caption{Leading order effective parton density of the photon as a function
of $Q^2$ at $P_T^2 = 85$~GeV$^2$.}
\label{fig:H1_epdf}
\end{figure}

H1 present the effective parton density of the photon, $\tilde{f}_\gamma$, as a 
function of $Q^2$ in Fig.~\ref{fig:H1_epdf}~\cite{J.Cvach}.
The data are consistent with parton densities which fall logarithmically with
$Q^2$ and are inconsistent with a pure vector meson dominance ansatz for the
photon's structure.

\section{JET SUBSTRUCTURE IN PHOTON-INDUCED COLLISIONS}

The measurement of jet substructure in photon-induced collisions has been used
to study universal properties of fragmentation.

\begin{figure}[h!]
\epsfig{file=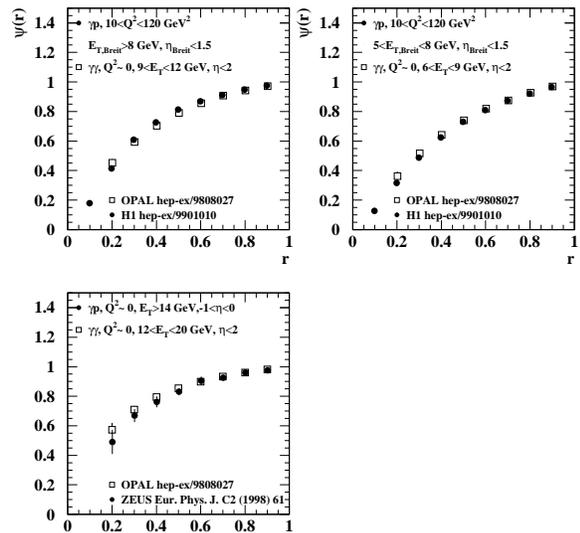,height=7.cm}
\vspace{-1.cm}
\caption{Jet shapes in dijet events measured by OPAL, H1 and ZEUS.}
\label{fig:shapes}
\end{figure}

Jet shapes measured in deep inelastic scattering and in $\gamma \gamma$
collisions are compared in the top row of
Fig.~\ref{fig:shapes}~\cite{S.Soeldner-Rembold}.
The comparison is made for measurements of similar $E_T^{\mb{jet}}$ and a
universality of fragmentation in jets is observed.  In the bottom plot
of this figure a comparison is made between jets in $\gamma \gamma$ 
collisions and in $\gamma p$ collisions.  The jets from the $\gamma p$
collisions are narrower, but this could be due to the slightly larger
$E_T^{\mb{jet}}$ of the $\gamma p$ measurement.

\begin{figure}[h!]
\centerline{\epsfig{file=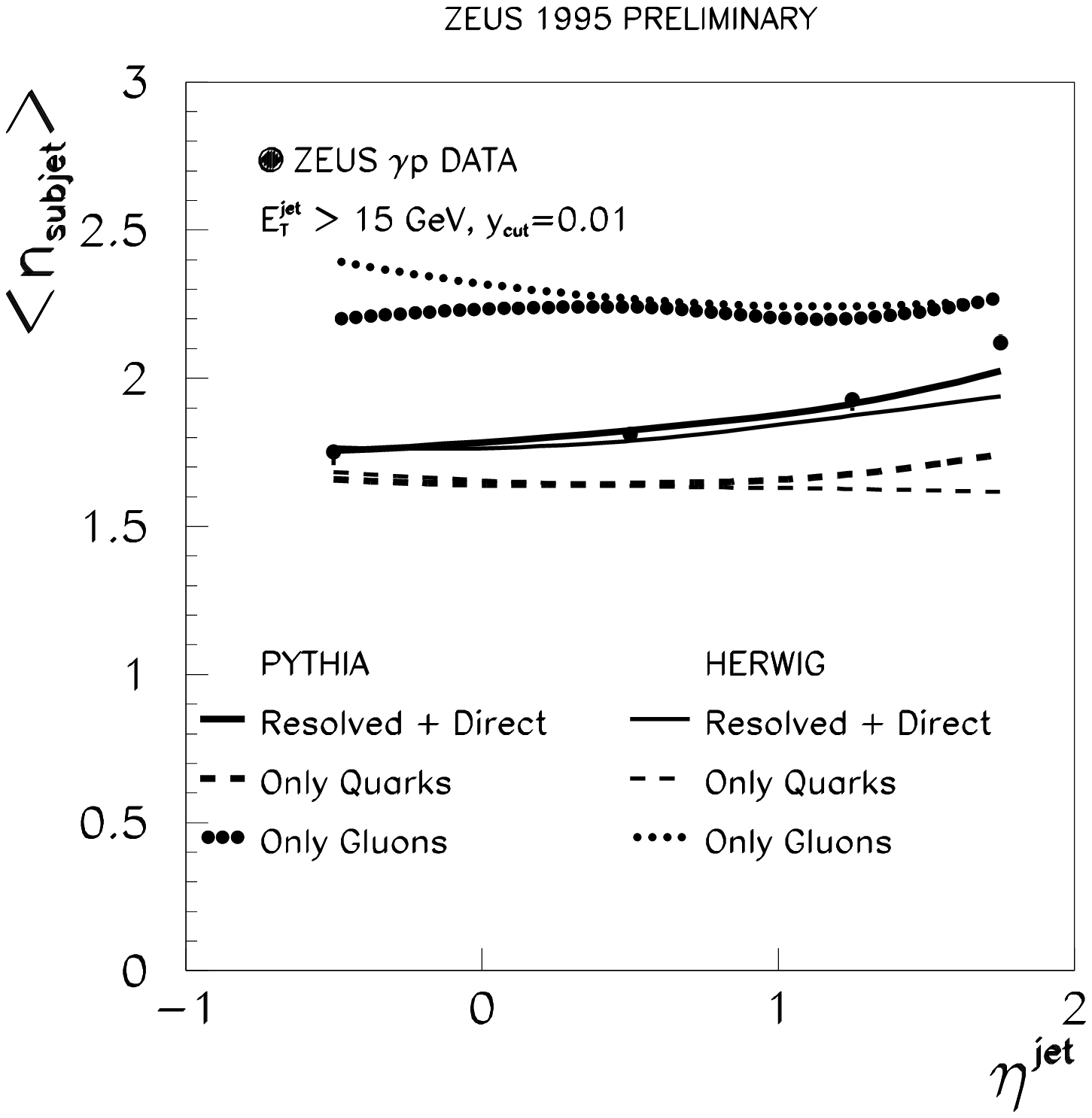,height=7cm,width=7cm}}
\vspace{-1.cm}
\caption{Number of subjets at $y_{\mb{cut}}=0.01$ as a function of
$\eta^{\mb{jet}}$.}
\label{fig:subjets}
\end{figure}

Using a clustering algorithm based on the relative transverse momentum of
particles, it is possible to resolve subjets within jets in a well-defined
manner.  The number of subjets will depend upon the value of the resolution
parameter, $y_{\mb{cut}}$, with which one looks into the jet.  In 
Fig.~\ref{fig:subjets} the mean number of subjets at resolution parameter
$y_{\mb{cut}} = 0.01$ is shown as a function of jet 
pseudorapidity~\cite{K.Umemori}. 
It is found that $\langle n_{\mb{subjet}}\rangle$ increases as expected 
from the  predominance of gluon jets in the forward region.

\section{INCLUSIVE PHOTONS IN $p \bar{p}$ COLLISIONS}

\begin{figure}[hbt]
\centerline{\psfig{figure=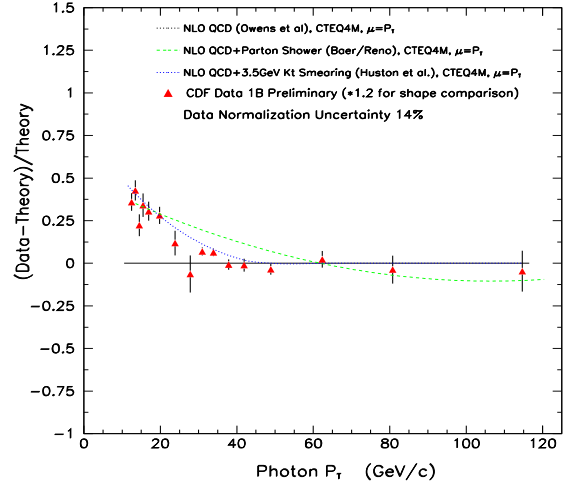,width=78mm,height=65mm}}
\vspace{-10mm}
\caption{The CDF measurement of the inclusive photon $p_T$ spectrum.}
\label{f:incPhoton}
\end{figure}

Inclusive photon results from the Tevatron were presented, where the 
CDF and D$\O$ data were seen to be consistent with each 
other~\cite{b:incPhoton}. The CDF data
reach lower photon $p_T$ values where the cross section  
tends to be higher than theory calculations. This shape difference is
difficult to explain with current NLO QCD calculations. 
Best fits are obtained with the $k_T$ smearing procedure used to explain 
the E706 photon data. A $k_T$ smearing of about 3.5 GeV is needed to explain 
the data. The result is shown in Fig.~\ref{f:incPhoton}.
In the same talk a high statistics measurement of photon-muon production 
was also presented, which compares well with NLO QCD.

\section{JET RESULTS FROM THE TEVATRON}

Results were presented on subjet multiplicity in quark and gluon jets
at D$\O$~\cite{b:subjets}. Jets are identified using the $k_T$ jet 
algorithm, which 
has the advantage of being IR-safe and allows a more direct comparison 
between theory and measurements. The gluon jet fraction was determined
at $\sqrt{s} = 1800$ GeV and $\sqrt{s} = 620$ GeV. The results indicate that
there are more subjets at $\sqrt{s} = 1800$ GeV as well as more subjets 
in gluon jets.

\begin{figure}[hbt]
\psfig{figure=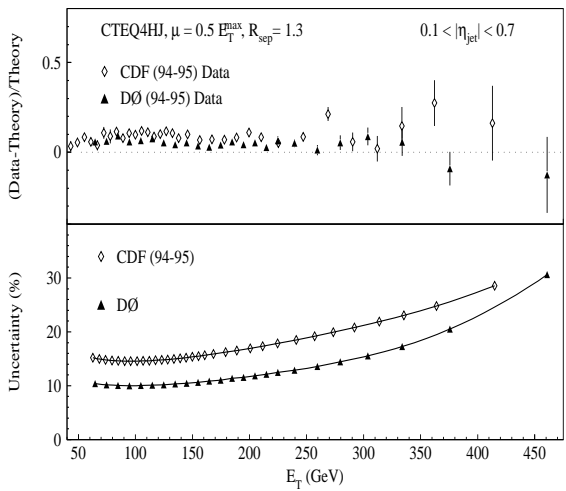,width=75mm,height=65mm}
\vspace{-10mm}
\caption{The CDF and D$\O$ measurements of the inclusive jet cross section.}
\label{f:incJet}
\end{figure}

In the joint session with the structure function working group,
both CDF and D$\O$ presented results on the inclusive 
jet cross section~\cite{b:d0IncJet}~\cite{b:cdfIncJet}.
When the CDF run IA results were published, they showed an excess
of events at high $E_T$ with respect to QCD expectations using particular
parton density functions. This excess generated a lot of interest, and 
explanations ranged from quark substructure to modified parton density
functions. The preliminary CDF measurement from the IB run is based on 
an integrated luminosity of 87.7 pb$^{-1}$ and is in agreement 
with the Run IA measurement. D$\O$ presented recently published results 
which are consistent with QCD predictions.
Improved energy calibrations at D$\O$ allowed them to reduce the 
systematic errors to 10\% at low $E_T$ and to about 30\% at high $E_T$.
A comparison of CDF data with D$\O$ data is shown in 
Fig.~\ref{f:incJet}, where the results are seen to be consistent. 
It appears that the behaviour at high $E_T$ can be accommodated 
by enhancing the gluon at high $x$ as is done in the CTEQ PDFs.
A more sensitive search for quark substructure can be conducted 
using either the dijet mass distribution or dijet angular distribution 
and will be discussed later.

A consistency check of $\alpha_s$ from jet data was presented by 
CDF~\cite{b:cdfIncJet}.
The technique extracts $\alpha_s$ from a third-order equation where
the coefficients are calculated assuming a particular PDF and 
value of $\alpha_s$. By varying  $\alpha_s$ one can check for a 
consistent solution where the extracted  $\alpha_s$ equals the input
value. The method depends on the choice of PDF since 
different PDFs result in a different $\alpha_s$.
The results show the running of $\alpha_s$ in one experiment and yield
a result consistent with measured $\alpha_s$ from other experiments.

\begin{figure}[hbt]
\psfig{figure=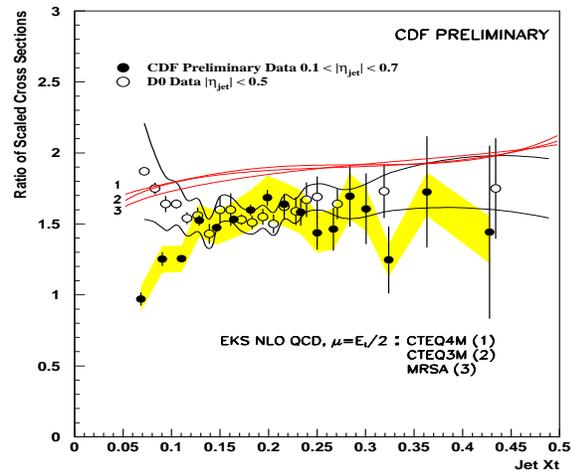,width=78mm,height=65mm}
\vspace{-10mm}
\caption{Ratio of the scaled cross section for centre-of-mass energies of 
1800 and 630 GeV as a function of $x_T$ as measured by CDF and D$\O$.}
\label{f:xtScaling}
\end{figure}

Both CDF and D$\O$ presented the ratio of the scaled cross section 
for a centre-of-mass energy of 630 and 1800 GeV 
as a function of $x_T$~\cite{b:d0IncJet}~\cite{b:cdfIncJet}. 
The ratio allows a reduction of the uncertainty due
to theory and experiment. Above values of $x_T = 0.1$  the CDF and 
D$\O$ measurements agree, while at lower $x_T$ values the measurements
diverge.  The D$\O$ data tend to higher ratio values 
while the CDF results tend to lower values as is shown in 
Fig.~\ref{f:xtScaling}.

D$\O$ also presented the rapidity dependence of the inclusive jet
cross section for three different $\eta$ bins up to 
$\eta < 1.5$~\cite{b:d0IncJet}.
Results were in good agreement with NLO QCD calculations.

\begin{figure}[hbt]
\psfig{figure=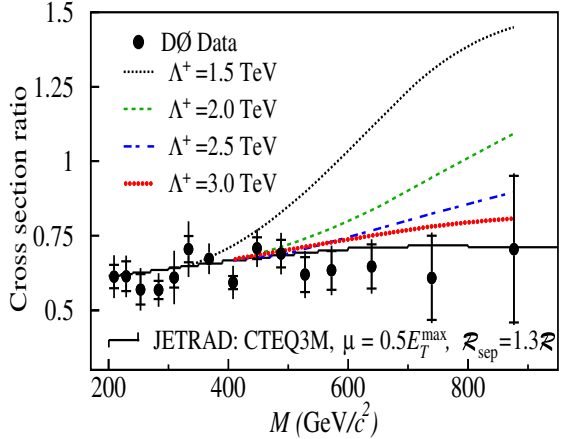,width=75mm,height=60mm}
\vspace{-10mm}
\caption{Limits placed on compositeness models using the D$\O$ dijet mass data}
\label{f:d0DijetmassLimits}
\end{figure}

Both CDF and D$\O$ presented measurements on the dijet mass 
distribution~\cite{b:d0dijet}~\cite{b:cdfdijet}. Results from the two 
experiments are in good agreement
in both shape and normalization. D$\O$ has used the measurement to
place limits on quark compositeness as is shown in 
Fig~\ref{f:d0DijetmassLimits}.
Dijet angular distributions provide a sensitive test of new physics
and have the advantage that the distributions are less sensitive to 
the energy measurement uncertainty. Results were presented from CDF
and used to place limits on quark compositeness.

D$\O$ presents the differential dijet cross section 
separately for opposite-side jets and same-side 
jets~\cite{b:d0dijet}.
Both jets are required to sit in the same $\eta$ bin. Results were
compared to the JETRAD calculation using different PDFs. 
Dijet differential cross sections from CDF were shown where the central
jet was used to measure the $E_T$ of
the event~\cite{b:cdfdijet}. A second jet is 
allowed to fall 
in one of four $\eta$ bins. A quantitative comparison of different PDFs
is under way. The differential dijet measurement covers a plane in 
the $x$--$Q^2$ space, making it more sensitive to the shape of the cross 
section determined by different PDFs. The data will provide a useful 
input to QCD fits in order to determine refined PDFs.
\begin{figure}[hbt]
\psfig{figure=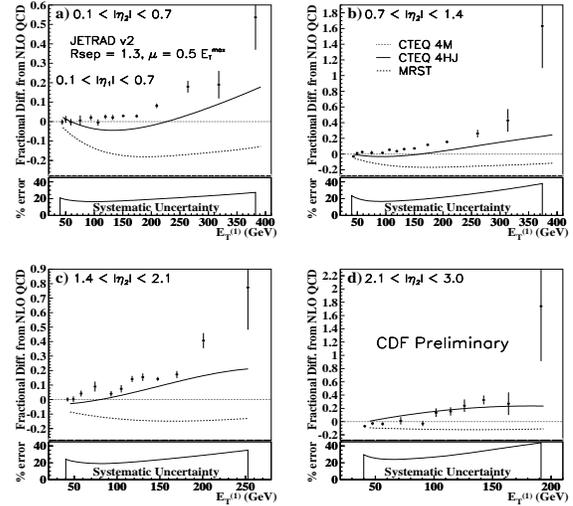,width=78mm,height=70mm}
\vspace{-10mm}
\caption{Triple differential dijet cross section.}
\label{f:dijetXsec}
\end{figure}

\section{CONCLUSIONS}

At this meeting many beautiful, high-precision experimental results were
presented and compared with theoretical predictions.
QCD has been clearly established as a successful theory for the 
description of hard scattering processes.
It is now increasingly important to better understand 
soft non-perturbative phenomena and processes where
more than one hard scale plays a part.
The further development of power corrections and resummed calculations
as well as the calculation of QCD cross sections at next-to-next-to-leading 
order is under way.
This program will eventually lead to an even more stringent comparison 
of theory and experiment.
The development of these concepts should benefit greatly from
continued close communication between theorists and experimentalists.

\end{document}